\documentclass[a4paper,11pt]{article}

\pdfoutput=1

\usepackage[a4paper,left=2.73cm,right=2.7cm,top=3cm,bottom=3.5cm]{geometry}

\usepackage[T1]{fontenc} 
\usepackage{graphicx}
\usepackage{epsfig}
\usepackage{amssymb}
\usepackage{amsfonts}
\usepackage{dsfont}
\usepackage{amsmath,euscript,array,mathrsfs,nccmath}
\usepackage{multirow}
\usepackage{bbold,bm,bbm}
\usepackage{epsf}
\usepackage{slashed}
\usepackage{comment}
\usepackage{colortbl}
\usepackage{lmodern}
\usepackage[utf8]{inputenc}
\usepackage[noadjust]{cite}
\usepackage{changepage}

\usepackage{jheppub}

\usepackage{caption,subcaption,wrapfig}
\usepackage[colorlinks=true,linktocpage=true,linkcolor=blue,citecolor=blue]{hyperref}

\usepackage{tikz}
\usetikzlibrary{decorations.markings,decorations.pathmorphing}

\DeclareMathOperator{\arccsc}{arccsc}

\newcommand{\be}{\begin{equation}} \newcommand{\ee}{\end{equation}}
\newcommand{\bea}{\begin{eqnarray}} \newcommand{\eea}{\end{eqnarray}}

\newcommand{\dd}{\text{d}}

\newcommand{\at}{\tilde{a}_1}
\newcommand{\ah}{\hat{a}_1}

\newcommand{\hIR}{ h_{\text{\tiny IR}}}

\newcommand{\UUU}{\mathbf{\mathcal{U}}}
\newcommand{\uc}{\varrho}
\newcommand{\nfour}{n_4}
\newcommand{\nfive}{n_5}
\newcommand{\nchone}{a_1}
\newcommand{\nchtwo}{a_2}

\newcommand\smallB{
  \mathchoice
    {{\scriptstyle\mathcal{B}}}
    {{\scriptstyle\mathcal{B}}}
    {{\scriptscriptstyle\mathcal{B}}}
    {\scalebox{.7}{$\scriptscriptstyle\mathcal{B}$}}
  }

  \newcommand\smallM{
  \mathchoice
    {{\scriptstyle\mathcal{M}}}
    {{\scriptstyle\mathcal{M}}}
    {{\scriptscriptstyle\mathcal{M}}}
    {\scalebox{.7}{$\scriptscriptstyle\mathcal{M}$}}
  }

\newcommand{\UU}{{\rm U}}
\newcommand{\SU}{{\rm SU}}

\newcommand{\newsec}[1]{\section{#1}}

\newcommand{\B}{\mathds{B}_8}
\newcommand{\VV}{\overline{\varsigma}}

\newcommand{\PreserveBackslash}[1]{\let\temp=\\#1\let\\=\temp}
\newcolumntype{C}[1]{>{\PreserveBackslash\centering}p{#1}}
\newcolumntype{R}[1]{>{\PreserveBackslash\raggedleft}p{#1}}
\newcolumntype{L}[1]{>{\PreserveBackslash\raggedright}p{#1}}

\begin{document}

	\begin{titlepage}
		
		\thispagestyle{empty}
		
		\begin{flushright}
			\hfill{}
		\end{flushright}
		
		\vspace{40pt}  
		
		\begin{center}

  \begin{adjustwidth}{-15mm}{-15mm}
  \begin{center}
      {\LARGE \textbf{Spontaneous breaking of baryon symmetry \\[6pt]  in strongly coupled 
      three-dimensional theories}}
  \end{center}
\end{adjustwidth} 
			
			\vspace{40pt}
			
			{\large \bf Ant\'on F. Faedo,$^{1,\,2}$   Carlos Hoyos,$^{1,\,2}$   \\ [1mm]
				
				and Javier G. Subils$^{3}$
			}

			\vspace{25pt}

			{\normalsize  $^{1}$ \textit{Departamento de F\'{i}sica, Universidad de Oviedo, \\ Calle Leopoldo Calvo Sotelo 18, ES-33007, Oviedo, Spain.}}\\
			\vspace{15pt}
			{ $^{2}$ \textit{Instituto Universitario de Ciencias y Tecnolog\'{\i}as Espaciales de Asturias (ICTEA), \\ Calle de la Independencia 13, ES-33004, Oviedo, Spain.}}\\
			\vspace{15pt}
			{ $^{3}$\textit{ Institute for Theoretical Physics, Utrecht University, \\ 3584 CC Utrecht, The Netherlands.}}\\
			\vspace{15pt}

			\vspace{60pt}
			\textbf{Abstract}
		\end{center} 
We show that baryon number symmetry is spontaneously broken in a class of three-dimensional, ${\cal N}=1$ supersymmetric theories with a discrete mass spectrum. These models serve as lower-dimensional, less-supersymmetric analogs of the Klebanov-Strassler solution, sharing properties such as the presence of a cascade. The spontaneous symmetry breaking is evidenced by the appearance of a Goldstone mode, which corresponds to a vector fluctuation in the gravity dual.
\end{titlepage}

\newpage

\tableofcontents

\newsec{Introduction}
\label{sec:intro}
Most non-perturbative tools in the supersymmetric kit fail to apply to theories with ${\cal N}=1$ supersymmetry in 2+1 dimensions, which in contrast to their more supersymmetric cousins are not protected from quantum corrections. For this reason, since the early work of Witten \cite{Witten:1999ds}, there have been few studies of their properties. However, more understanding about the vacuum structure has been gained in recent years \cite{Gomis:2017ixy,Bashmakov:2018wts,Benini:2018umh,Benini:2018bhk,Bashmakov:2021rci,Armoni:2023ohe}. 

The gauge/gravity duality can be used to further the study of ${\cal N}=1$ theories in the large-$N$ limit. A motivation is that they have string theory realizations, and thus the two sides of the duality are known to exist. Furthermore, even though their dynamics might be closer to those of non-supersymmetric theories, their stability is still protected, while most non-supersymmetric solutions are unstable or suspected to be, at any rate, a question that might be hard to dilucidate. Thus, holographic duals of ${\cal N}=1$ theories serve both to study the properties of their strongly coupled field theory counterparts and as proxies for theories, hopefully closer to the real world.

We will be concerned here with a family of supergravity solutions dual to supersymmetric ${\cal N}=1$ theories with a discrete mass spectrum\footnote{As explained in \cite{Faedo:2017fbv}, only one type of solution in this family corresponds to a confining theory according to the criterion of area law for Wilson loops.}  \cite{Cvetic:2001bw,Herzog:2002ss,Faedo:2017fbv,Cvetic:2001pga}. Although the field theory dual is not exactly known, it is expected to be a deformation of the field theory living at the intersection of D2, D4 and D6 branes. The gauge group is $\UU(N_1)\times \UU(N_2)$ or $\SU(N_1)\times \SU(N_2)$, with Chern-Simons level $k$ and matter in the bifundamental representation. The ranks and the level depend on the number of each type of brane at the intersection. As explained in \cite{Faedo:2023nuc}, following the reasoning of \cite{Bergman:2020ifi}, whether the group is unitary or special unitary depends on the boundary conditions of the fields in the supergravity solution. In the following we will be concerned only with the special unitary case, which is the one enjoying a baryonic symmetry, as we explain later.

The phase diagrams for these theories were studied in \cite{Elander:2020rgv,Faedo:2022lxd,Faedo:2023nuc}. In particular, in \cite{Faedo:2023nuc} it was proposed using thermodynamic arguments that, in the confining phase of the $k=0$ theory, the baryon symmetry should be spontaneously broken.\footnote{Note that the Vafa-Witten theorem \cite{Vafa:1983tf,Vafa:1984xh} can be evaded, due to the presence of scalars charged under baryon symmetry.} This would suggest that the supergravity solution is dual to a `non-Abelian' vacua with spontaneously broken global symmetry similar to those predicted in purely field theoretical analysis of other ${\cal N}=1$ theories \cite{Bashmakov:2018wts,Bashmakov:2021rci}. Our goal in this work is to put this on firm ground by explicitly finding the Goldstone mode associated to the spontaneous breaking of baryon symmetry. Additionally, we will show that the Goldstone mode is also present when the Chern-Simons level is non-zero, so we expect baryon symmetry to be spontaneously broken in the ground state of any of the theories with duals within the family of supergravity solutions we study that have a massive spectrum.

The outline of the paper is as follows: in section \ref{sec:setup} we introduce the ${\cal N}=1$ theories with their gravity duals and show the equations of vector modes and ansatz that capture the Goldstone mode. In sections \ref{sec:conf} and \ref{sec:nonconf} we find the solution of the vector modes dual to the Goldstone mode in the dual of a confining theory and of a non-confining theory with a massive spectrum, respectively. Finally, we discuss the results and comment on possible future directions in section \ref{sec:discuss}.

\newsec{Vector equations of motion and fluctuation ansatz} 
\label{sec:setup}

The special unitary ${\cal N}=1$ gauge theories that we study have a global baryonic symmetry, ${\rm U}(1)_{\smallB}$, that acts on multiplets in the (anti)bifundamental representation. As mentioned earlier, it was argued in \cite{Faedo:2023nuc} that the global baryonic symmetry is spontaneously broken in the confining phase of the $k=0$ theory. As a consequence, the spectrum of fluctuations should contain a massless mode corresponding to the associated Goldstone boson, analogous to the one found in \cite{Aharony:2000pp,Gubser:2004qj} in the Klebanov--Strassler background. Similarly to \cite{Gubser:2004qj}, the Goldstone mode should appear as a perturbation of the appropriate vector field in the reduced four-dimensional supergravity theory around the confining geometry.\footnote{Indeed, no massless modes were found in the perturbations of the scalar sector of the reduced four-dimensional model \cite{Elander:2018gte}.} We will see that the Goldstone is actually present even at non-zero Chern-Simons level. This is in contrast to other related models, see \cite{Choi:2018tuh}.

To identify this perturbation, we will work directly with a consistent truncation in four dimensions. The details of the reduction and the notation we use can be found in \cite{Faedo:2022lxd}. The four-dimensional supergravity theory contains a massless vector $a_1$, together with two massive vectors, $\tilde{a}_1$ and $\hat{a}_1$, and a massive two-form $b_2$. In addition, there are six scalars $\Phi,U,V,b_X,b_J,a_J$ and an axion $a_X$.

The massless vector is of particular interest since it is dual to a conserved current. It descends from the ten-dimensional RR one-form as $C_1=a_1$, so in particular it couples electrically to D0-branes through their Wess--Zumino term
 \begin{align}
 T_{\text{\tiny D0}}\int C_1=T_{\text{\tiny D0}}\int a_1\,.
 \end{align}
When the field $a_1$ obeys Dirichlet boundary conditions, D0-branes can end at the boundary and are dual to local monopole operators in the three-dimensional gauge theory \cite{Faedo:2023nuc}. This leads us to identify $a_1$ with the monopole number current associated with the magnetic ${\rm U}(1)_{\smallM}$ symmetry in the $\UU(N_1)\times\UU(N_2)$ theory. On the other hand, the baryonic symmetry ${\rm U}(1)_{\smallB}$ is only present in the ${\rm SU}(N_1)\times {\rm SU}(N_2)$ theory, which corresponds to Neumann boundary conditions for $a_1$. In that case, a D6-brane wrapping the entire internal manifold ${\cal M}_6$ can  end at the boundary. This corresponds to a local baryon operator in the field theory dual \cite{Faedo:2023nuc}. The gauge field dual to the baryon number current is thus the one electrically coupled to this brane, that is, $A_1\sim\int_{{\smallM}_6} C_7$.

\subsection{Vector equations}

To lighten the notation, it will be convenient to introduce the field strengths
\begin{align}\label{eq:fieldstrength}
\tilde{f}_2=\dd \at+b_X\,\dd a_1+Q_k \,b_2\,,\qquad\qquad\hat{f}_2=\dd \ah+b_J\,\dd a_1-Q_k\, b_2\,,    
\end{align}
and a covariant derivative for the axion $Da_X=\dd a_X-\at-\ah$. We also define the following combinations of scalars  
\begin{align}
    *f_4&=-16e^{-6U-12V-\Phi/2}\left[4a_J\left(b_X+b_J\right)+2q_c\left(b_X-b_J\right)+Q_kb_J\left(b_J-2b_X\right)\right]\,,\nonumber\\[2mm]
    f_0^J&=2a_J-Q_kb_J+q_c\,,\qquad\qquad f_0^X=2a_J+Q_k(b_J-b_X)-q_c\,.
\end{align}
The constants $q_c$ and $Q_k$ are Page charges for D4 and D6-branes, respectively, related with gauge theory parameters as 
\begin{align}
    q_c\,=\,\frac{3\pi\ell_s^3g_s}{4}\,\left(M-\frac{k}{2}\right)\,,\qquad\qquad Q_k\,=\,\frac{\ell_sg_s}{2}\,k\,,
\end{align}
where $g_s$ and $\ell_s$ are the string coupling and length respectively, $k$ is the Chern-Simons level and $M=|N_1-N_2|$ the difference between the ranks of the two gauge groups in the dual theory.

With these definitions, the dynamics of the vector fields are described by the following equations of motion for the massless vector
\begin{align}\label{eq:eoma1}
\dd\left(e^{2U+4V+3\Phi/2}*\dd a_1\right)+e^{6U+12V+\Phi/2}\left(*f_4\right)\dd b_2+8e^{2U+\Phi/2}\dd b_J\wedge*\hat{f}_2&\nonumber\\[2mm]
+16e^{-2U+4V+\Phi/2}\dd b_X\wedge*\tilde{f}_2+64\left(b_X+b_J\right)e^{-2U-4V+\Phi/2}*Da_X&=0\,,\end{align}
massive vectors
\begin{align}\label{eq:eoma1ha1t}
&\dd\left(e^{2U+\Phi/2}*\hat{f}_2\right)-8e^{-2U-4V+\Phi/2}*Da_X+4f_0^X\dd b_2+4\dd b_J\wedge\tilde{f}_2+4\dd b_X\wedge\hat{f}_2=0\,,\nonumber\\[2mm] 
&\dd\left(e^{-2U+4V+\Phi/2}*\tilde{f}_2\right)-4e^{-2U-4V+\Phi/2}*Da_X+2f_0^J\dd b_2+2\dd b_J\wedge\hat{f}_2=0\,,
\end{align}
and two-form
\begin{align}\label{eq:eomb2}
\dd\left(e^{4U+8V-\Phi}*\dd b_2\right)-e^{6U+12V+\Phi/2}\left(*f_4\right)\dd a_1-16Q_ke^{-2U+4V+\Phi/2}*\tilde{f}_2&\nonumber\\[2mm]
+8Q_ke^{2U+\Phi/2}*\hat{f}_2-32f_0^J\,\tilde{f}_2-32f_0^X\,\hat{f}_2-64\,Da_X\wedge\dd a_J&=0\,.
\end{align}
The axion covariant derivative $D a_X$ is responsible for a mass term for the combination $\at+\ah$, since the axion $a_X$ can be taken to vanish. In all these expressions, the scalars $U$, $V$, $\Phi$, $b_J$, $b_X$ and $a_J$ are specified by choosing a background, given that for our purposes of identifying a massless mode in the spectrum it is sufficient to consider linearized equations for the fluctuations. 
It can be shown that the exterior derivative of \eqref{eq:eomb2} is equivalent to a combination of the equations of motion for the massive vectors, reflecting the fact that the two-form becomes massive by eating a vector. This means that the three equations in \eqref{eq:eoma1ha1t} and \eqref{eq:eomb2} are not independent.

The duality relation $*_{10}\, F_2=F_8$ between the field strengths of $C_1$ and $C_7$ translates in four dimensions into the fact that the massless vectors $a_1$ and $A_1$, corresponding to ${\rm U}(1)_{\smallM}$ and ${\rm U}(1)_{\smallB}$ respectively, are electromagnetic duals. This can be seen by noting that the left hand side of \eqref{eq:eoma1} is a total derivative and therefore the equation can be solved by 
\begin{align}\label{eq:duality}
&e^{2U+4V+3\Phi/2}*\dd a_1+16\,b_X\,e^{-2U+4V+\Phi/2}*\tilde{f}_2+8\,b_J\,e^{2U+\Phi/2}*\hat{f}_2\nonumber\\[3mm]
&-16\,b_J^2\,b_X\,\dd a_1+16\,b_J^2\,\tilde{f}_2+32\,b_X\,b_J\,\hat{f}_2=\dd A_1\,,
\end{align}
where $A_1$ is precisely the vector contained in $C_7$ with the correct coupling to the baryonic D6-branes. The Goldstone mode we are looking for should then be a normalizable, regular perturbation of this massless vector. 

For technical reasons, we solve directly equations \eqref{eq:eoma1}-\eqref{eq:eomb2}. Then we interpret the results in terms of the baryonic vector $A_1$ using the convoluted duality condition \eqref{eq:duality}. Notice in particular that electric charge and magnetic field are interchanged through the duality. 

\subsection{Ansatz for the Goldstone mode}\label{sec:ansatz}

A consistent ansatz to look for a massless mode in the spectrum of vector fluctuations is as follows. Let $a(t,x_1,x_2)$ be a harmonic function at the boundary, that is, it verifies 
\begin{align}\label{eq:harmonic}
    \dd *_3\dd a=0\,,
\end{align}
with $*_3$ the Hodge duality operation in three-dimensional Minkowski space $\dd x_{1,2}^2$. In terms of this function, the massless vector is taken to be 
\begin{align}\label{eq:ansatz_a1}
    \dd a_1=\dd\left(\alpha\,\dd a\right)+\sigma_{\smallM}*_3\dd a\,,
\end{align}
where $\alpha$ is a function of the holographic radial coordinate and $\sigma_{\smallM}$ is a constant. The mass of this fluctuation is clearly vanishing due to the condition \eqref{eq:harmonic}.

The massive vectors and two-form couple to this vector, so we need to solve also for their profiles. These can be written in terms of the same harmonic function and new functions of the holographic radial coordinate $\tilde\alpha$, $\hat\alpha$, $\beta$ and $\mathcal{B}$, as
\begin{align}
\tilde{a}_1&=\tilde{\alpha}\,\dd a\,,&\quad\quad\quad& \hat{a}_1=\hat{\alpha}\,\dd a\,,\quad\quad\quad &b_2=\beta\,*_3\dd a+\dd\left(\mathcal{B}\,\dd a\right)\,.
\end{align}
The second term in $b_2$ does not contribute to $\dd b_2$, but it does enter the equations when $Q_k\ne0$ through the field strength of the vectors \eqref{eq:fieldstrength} and, moreover, plays a role in the gauge fixing procedure. In this case the vectors, being massive, do not admit a term proportional to $*_3\dd a$. The solutions that correspond to a Goldstone mode will be both regular in the interior and normalizable at the boundary. In this model, this means that they vanish at the boundary.

Finally, the vector dual to the baryon number symmetry admits an analogous ansatz 
\begin{align}\label{eq:A1_ansatz}
\dd A_1=\dd\left(\mathcal{A}\,\dd a\right)+\sigma_{\smallB}*_3\dd a
\end{align}
for another function of the holographic radial coordinate $\mathcal{A}$ and a constant $\sigma_{\smallB}$. The Goldstone thus corresponds to a profile for $\mathcal{A}$ with the correct boundary conditions. The parameter $\sigma_{\smallB}$ introduces a background magnetic field for the baryon currents, so it should vanish for the solution that corresponds to a Goldstone mode. In addition, the function $\mathcal{A}$ should be regular in the interior and vanish at the boundary. However, it is always possible to shift $\mathcal{A}$ by a constant, so the latter constraint only fixes this arbitrary constant. We will specify further the regularity conditions for all the fields below.

\section{Vector modes in the confining theory}\label{sec:conf}

In this section, we solve the vector equations \eqref{eq:eoma1}-\eqref{eq:eomb2} in a confining background with the appropriate boundary conditions to interpret the solution as a massless excitation in the dual special unitary gauge theory. 

Within the family of solutions to the four-dimensional supergravity equations, only those dual to a theory with vanishing Chern--Simons level $k=0$ are truly confining, while those with $k\neq 0$ have massive excitations but do not confine \cite{Faedo:2017fbv}. Therefore, we will take $Q_k=0$ in the equations.
In this case, the system can be slightly simplified by observing that, for vanishing $Q_k$, the left-hand side of \eqref{eq:eomb2} is a total derivative\footnote{For more details, see Appendix B in \cite{Faedo:2022lxd}. Recall also that one can fix the axion $a_X$ in $Da_X$ to any constant value, in particular zero.}
\begin{align}
    \dd\left[e^{4U+8V-\Phi}*\dd b_2+64a_J\,Da_X+32q_c\left(\ah-\at\right)\right]=0\,.
\end{align}
This allows us to express the derivative of the two-form in terms of the massive vectors as
\begin{align}\label{eq:derivative_two_form}
    \dd b_2=32\,e^{-4U-8V+\Phi}*\left[q_c\left(\at-\ah\right)+2a_J(\at+\ah)\right]\,.
\end{align}
Given that for $Q_k=0$ in the remaining system only the derivative of the two-form appears, we can substitute it by \eqref{eq:derivative_two_form}, reducing the number of equations to be solved. In particular, $\mathcal{B}$ does not appear in the system since it is pure gauge. 

\subsection{Confining background geometry}

Using the expressions for the reduction in \cite{Faedo:2022lxd}, the four-dimensional metric corresponding to the confining geometry reads 
\begin{align}\label{eq:background_1}
    \dd s_4^2=e^{2U+4V-\Phi/2}\left[h^{-1/2}\dd x_{1,2}^2+h^{1/2}\left(1-\frac{\rho_0^4}{\rho^4}\right)^{-1}\dd\rho^2\right]
\end{align}
where $\rho\in(\rho_0,\infty)$ is the radial coordinate, $h$ is the ten-dimensional warp factor and $\rho_0$ is a constant with dimensions of length that we identify with gauge theory parameters as \cite{Faedo:2022lxd}
\begin{align}
    \rho_0=\frac{3g_s\ell_sK\left(-1\right)}{2}\,\frac{M^2}{N}\,,
\end{align}
with $K(m)$ the complete elliptic integral of the first kind.\footnote{In this case $K(-1)={\sqrt{\pi}\,\Gamma(\frac{5}{4})}\big/{\Gamma(\frac{3}{4})}=1.311\dots$.} The dilaton and scalars originating from the ten-dimensional metric are given also in terms of the warp factor as 
\begin{align}\label{eq:background_2}
    e^{2U}=h^{3/8}\rho^2\left(1-\frac{\rho_0^4}{\rho^4}\right)\,,\qquad\qquad e^{2V}=h^{3/8}\rho^2\,,\qquad\qquad e^{\Phi}=h^{1/4}\,.
\end{align}
The scalars descending from the ten-dimensional fluxes, written as a function of the dimensionless radial coordinate $z=\rho/\rho_0$, are 
\begin{align}\label{eq:background_fluxes}
b_J&=\frac{2q_c}{3\rho_0}\left[\frac{z\sqrt{z^4-1}-\left(3z^4-1\right)\UUU(z)}{z^4-1}\right]\,,\nonumber\\[2mm]
b_X&=-\frac{2q_c}{3\rho_0}\left[\frac{z\sqrt{z^4-1}-\left(3z^4-1\right)\UUU(z)}{z^4}\right]\,,\\[2mm]
a_J&=\frac{q_c}{6}+\frac{2q_c\,\UUU(z)}{3z\sqrt{z^4-1}}\,,\nonumber
\end{align}
where the dimensionless function $\UUU$ is defined as 
\begin{align}
\UUU(z)\,=\,\int_1^z\left(\sigma^4-1\right)^{-1/2}\dd \sigma=K\left(-1\right)-F\left(\arccsc z|-1\right)\,,
\end{align}
with $F\left(\phi|m\right)$ the elliptic integral of the first kind. 
On the other hand, the warp factor cannot be found in closed form. It can be expressed in terms of the integral
\begin{align}\label{eq:background_3}
h=\frac{128\,q_c^2}{9\,\rho_0^6}\int_z^\infty\left[\frac{2-3\sigma^4}{\sigma^3\left(\sigma^4-1\right)^2}+\frac{\left(4-9\sigma^4+9\sigma^8\right)\UUU(\sigma)}{\sigma^4\left(\sigma^4-1\right)^{5/2}}+\frac{2\left(1-3\sigma^4\right)\UUU(\sigma)^2}{\sigma^5\left(\sigma^4-1\right)^3}\right]\dd\sigma.
\end{align}
In ten dimensions the geometry ends smoothly at $\rho=\rho_0$. Note that the warp factor is a finite quantity in the IR, since $h(z=1)= {128\,q_c^2\hIR}/({9\,\rho_0^6})$ with $\hIR \simeq 0.8554$.

\subsection{Vector mode equations}

The electromagnetic duality condition relates the baryonic quantities $\mathcal{A}$ and $\sigma_{\smallB}$ to $\alpha$ and $\sigma_{\smallM}$. In fact, substituting the ansatz for the vectors in \eqref{eq:duality} and particularizing to the confining background we obtain two equations. The first one reads 
\begin{align}\label{eq:monopole_charge}
h\left(\rho^4-\rho_0^4\right)^{3/2}\alpha'&+\frac{16b_X\rho^4}{\left(\rho^4-\rho_0^4\right)^{1/2}}\left(\tilde{\alpha}'+b_X\alpha'\right)\nonumber\\[2mm]&+\frac{8b_J\left(\rho^4-\rho_0^4\right)^{3/2}}{\rho^4}\left(\hat{\alpha}'+b_J\alpha'\right)-32\sigma_{\smallM} b_J^2b_X=-\sigma_{\smallB}\,,
\end{align}
where prime denotes derivative with respect to the radial coordinate $\rho$ and the warp factor and other scalars are given by \eqref{eq:background_fluxes} and \eqref{eq:background_3} respectively.

The second equation obtained from \eqref{eq:duality} fixes the derivative of $\mathcal{A}$ in terms of the other vectors as 
\begin{align}\label{eq:A1_profile}
\mathcal{A}'= \frac{\sigma_{\smallM}h\left[\left(\rho^4-\rho_0^4\right)\left(h\rho^4+8b_J^2\right)+16\rho^8b_X^2\right]}{\left(\rho^4-\rho_0^4\right)^{3/2}}+16b_J\left(2b_X\hat{\alpha}'+b_J\tilde{\alpha}'+2b_Jb_X\alpha'\right)\,.  
\end{align}
On the other hand, substituting the ansatz for the Goldstone mode in \eqref{eq:eoma1ha1t}, particularized again to the confining solution, gives two second-order ordinary differential equations, for the radial functions $\tilde{\alpha}$ and $\hat{\alpha}$, that read
\begin{align}\label{eq:radial_eoms}
&\left[\frac{\rho^4}{\left(\rho^4-\rho_0^4\right)^{1/2}}\left(\tilde{\alpha}'+b_X\alpha'\right)\right]'-\frac{4\rho^2}{\left(\rho^4-\rho_0^4\right)^{1/2}}\left(\tilde{\alpha}+\hat{\alpha}\right)-\left(4a_J+2q_c\right)\beta'-2\sigma_{\smallM} b_J b_J'=0\nonumber\\[2mm]
&\left[\frac{\left(\rho^4-\rho_0^4\right)^{3/2}}{\rho^4}\left(\hat{\alpha}'+b_J\alpha'\right)\right]'-\frac{8\rho^2}{\left(\rho^4-\rho_0^4\right)^{1/2}}\left(\tilde{\alpha}+\hat{\alpha}\right)-\left(8a_J-4q_c\right)\beta'-4\sigma_{\smallM} \left(b_Jb_X\right)'=0
\end{align}
 The two-form radial profile $\beta(\rho)$ is given in terms of the massive vectors through \eqref{eq:derivative_two_form}, which in this particular ansatz is 
\begin{align}
    \beta'=\frac{32}{h\left(\rho^4-\rho_0^4\right)^{3/2}}\left[q_c\left(\tilde{\alpha}-\hat{\alpha}\right)+2a_J\left(\tilde{\alpha}+\hat{\alpha}\right)\right]\,.
\end{align}
Notice that in all these equations the massless vector radial profile, $\alpha$, only appears through derivatives. Thus, we can solve for $\alpha'$ in \eqref{eq:monopole_charge} and substitute in the rest of the system. In this way, we reduce the problem to two second-order ordinary differential equations \eqref{eq:radial_eoms} for the massive vectors. The resulting functions $\tilde{\alpha}$ and $\hat{\alpha}$ then determine the profile of the Goldstone mode, $\mathcal{A}$ in \eqref{eq:A1_ansatz}, through \eqref{eq:A1_profile}.

Unfortunately, the equations are still quite involved, preventing us from finding analytic expressions for the solution. For this reason, we resort to numerical methods.

\subsection{Vector mode solutions}

We examine first the asymptotic behaviour of the fields. 
The boundary is located at $\rho = \infty$. Demanding that the solutions are normalizable, the functions fall off as 
\begin{equation}\label{eq:expasions_conf_UV}
    \hat \alpha = \frac{8}{9} K(-1)^2 \varsigma \frac{\rho_0^3}{\rho}+\cdots  +\nu _4 \frac{\rho_0^6}{\rho^4}+
    \nu _5 \frac{\rho_0^7}{\rho^5}
    +\cdots\,,\qquad
    \tilde \alpha = -\frac{8}{9} K(-1)^2 \varsigma  \frac{\rho_0^3}{\rho} +\cdots\,,
\end{equation}
where we have defined the dimensionless variable $\varsigma = q_c^2 \sigma_{\smallM} /\rho_0^5$. Only the leading and subleading  coefficients that are not fixed by the equations are shown explicitly. In particular, there are two undetermined dimensionless parameters, $\nu_4$ and $\nu_5$. The normalizable solution for $\mathcal{A}$ has the following boundary expansion
\begin{equation}
    \mathcal{A} = q_c^2\left(-\frac{6656}{9} K(-1)^4 \varsigma\, \frac{\rho_0}{\rho}+\frac{1900544 K(-1)^3 \varsigma}{2835}\, \frac{\rho_0^2}{\rho^2} +\cdots\right).
\end{equation}

Similarly, demanding that the functions remain finite in the interior, one finds
\begin{equation}\label{eq:expasions_conf_IR}
        \hat\alpha = \rho_0^2\left(\frac{64 a_0 \left(3-4 \hIR\right)-3 \sigma_{\smallB}/(\rho_0^ 2q_c)}{256 \hIR} +a_1\tau + \cdots + a_3\tau^3+\cdots\right),\quad
		\tilde\alpha = a_0 \rho_0^ 2+\cdots\,
\end{equation}
where, $\tau = \sqrt{1-\rho_0/\rho\, }$. Again, only the coefficients that are not fixed by the equations of motion are shown. There are three undetermined parameters: $a_0, \, a_1,$ and $a_3$. However, when the solution is uplifted to ten dimensions, the massive vector $\tilde{a}_1$, whose profile is $\tilde\alpha$, accompanies a cycle that collapses to zero size in the IR \cite{Faedo:2022lxd}. Thus, in order to prevent a non-vanishing flux piercing a collapsing cycle, we have to set $a_0=0$. Furthermore, when substituting Eq.~\eqref{eq:expasions_conf_IR} into Eq.~\eqref{eq:monopole_charge} it is found that
\begin{equation}
    \alpha = \frac{\sigma_{\smallB}}{18432\hIR\rho_0^ 2q_c}\,\,\frac{1}{\tau^ 3}+\cdots\,,
\end{equation}
so regularity also requires $\sigma_{\smallB} = 0$. Recall that this is precisely the requirement for the absence of a background magnetic field for the baryon current. This eliminates all the divergent terms in $\alpha$ and completes all the regularity and normalizability conditions. 

\begin{figure}[t]\centering\includegraphics[width=.48\textwidth]{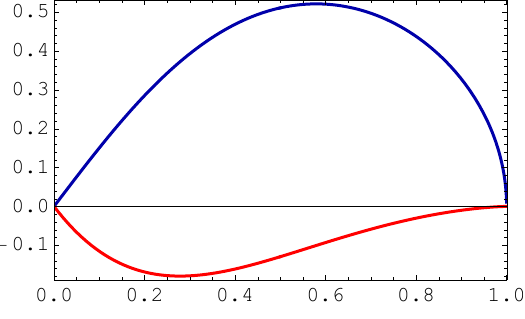}\hfill
	\put(-40,-5){$\rho_0/\rho$}
	\put(-180,130){$\tilde\alpha/\rho_0^{2}$,\, $\hat\alpha/\rho_0^{2}
    $}
 \hfill\includegraphics[width=.48\textwidth]{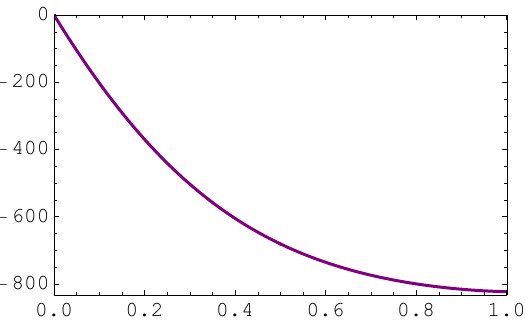}\put(-40,-5){$\rho_0/\rho$}
	\put(-180,130){$\mathcal{A}/q_c^2$}
	\caption{\small Left: numerical solution for $\hat\alpha$ (blue) and $\tilde\alpha$ (red) 
    that interpolates smoothly between the two asymptotic regimes \eqref{eq:expasions_conf_UV}, \eqref{eq:expasions_conf_IR}. The boundary is located at $\rho_0/\rho\to0$. Right: corresponding profile for $\mathcal{A}$ as given by \eqref{eq:A1_profile}. 
    }\label{fig.Goldstone_B8conf}
\end{figure}
The existence of a solution with the boundary conditions above, and therefore a Goldstone mode, requires matching numerically the boundary and interior expansions. This means finding an appropriate combination of $\varsigma$, $a_1$, $a_3$, $\nu_4$ and $\nu_5$ such that the solution integrated from the IR connects smoothly with the solution integrated from the UV.
However, note that the system is linear and homogeneous with respect to these variables: if a solution exists, it is always possible to rescale it to obtain another one. This allows us to fix $\varsigma=1$. Demanding continuity of $\tilde\alpha$, $\hat\alpha$ and their derivatives, we find a solution  for which \begin{equation}
	\nu_4 \simeq 2.17,\quad
	\nu_5 \simeq 3.41,\quad
	a_1 \simeq 1.09,\quad
	a_3 \simeq -0.43.
\end{equation}
This is plotted in the left panel of Fig.~\ref{fig.Goldstone_B8conf} in terms of the radial coordinate $\rho_0/\rho=z^{-1}$. Given these functions, it is straightforward to integrate Eq.~\eqref{eq:A1_profile} to find the profile for the baryonic vector, $\mathcal{A}$, which is shown in the right panel of the same figure.

\section{Vector modes in the non-confining theory}\label{sec:nonconf}

Let us now study the effect of including Chern--Simons interactions, which corresponds to $Q_k\neq 0 $. Although the theory still has a massive spectrum, a non-vanishing $Q_k$ immediately spoils the confining behavior, understood as an area law for Wilson loops, see \cite{Faedo:2017fbv}. Our aim is to understand the fate of the Goldstone mode when this drastic modification is introduced and whether the ground state still breaks any baryonic symmetry spontaneously. 

\subsection{Non-confining background geometry}

When $Q_k \neq 0$ there is actually a family of geometries labeled by the asymptotic value of the NS two-form on the two-cycle. This corresponds, in the dual field theory, to the asymptotic UV difference between the gauge couplings of both groups in the quiver. We will focus on a representative of this family, dubbed $\B$, which is the only one for which the entire solution is known in closed, analytic form. 
The metric in four dimensions is
\begin{align}
\dd s_4^2=e^{2U+4V-\Phi/2}\left[h^{-1/2}\dd x_{1,2}^2+h^{1/2}\dd r^2\right]\,.
\end{align}
The dilaton and scalars appearing in the uplifted metric are given in terms of the warp factor as 
\begin{equation}
   e^{2U}=h^{3/8}\,\frac{\left[r(r-2r_0)\right]^{3/2}}{r-r_0},\quad
   e^{2V}=h^{3/8}\,\left[r^3(r-2r_0)\right]^{1/2},\quad
   e^{\Phi}= h^{1/4}\,\frac{r(r-2r_0)}{(r-r_0)^2}\,,
\end{equation}
where the UV asymptotics fix the constant $r_0 = 2|Q_k|$ and the space ends smoothly\footnote{This family of geometries is regular in eleven dimensions but not in type IIA supergravity. For details of the uplift, see \cite{Faedo:2017fbv}.} at $r = 2r_0 = 4|Q_k|$.
Similarly to the previous case, several components of the eleven-dimensional flux are turned on to render the geometry regular, which introduces the four-dimensional scalars
\begin{subequations}
\begin{align}
b_J&=\frac{2q_c\left(r-2r_0\right)\left(-3r^3+r_0\,r^2+2\,r_0^2\,r-4r_0^3\right)}{15|Q_k|r^3(r-r_0)}\,,\\[3mm]
b_X&=\frac{2q_c(r-2r_0)^2(3r^3+2\,r_0\,r^2+r_0^2\,r-4r_0^3)}{15|Q_k|r^3(r-r_0)^2}\,,\\[3mm]
a_J&=\frac{q_c(-r^3+16\,r_0\,r^2+48\,r_0^2\,r-32\,r_0^3)}{30\,r^3}\,.
\end{align}
\end{subequations}
Finally, the warp factor can be found in analytic form and reads\footnote{This is the warp factor in ten dimensions, so it is not regular in the IR at $r=2r_0$. The regular, eleven-dimensional warp factor $H$ is related to this one as $H=\frac{r(r-2r_0)}{(r-r_0)^2}\,h$.}
\begin{align}
h=\frac{32q_c^2r_0(1323r^6+924r_0r^5+963r_0^2r^4+510r_0^3r^3-1340r_0^4r^2-4340r_0^5r+2800r_0^6)}{23625|Q_k|^2r^{10}(r-2r_0)}\,.
\end{align}

Even though the background is completely determined by these expressions, we were unable to solve the equations for the vector fluctuations analytically, so we proceed with a numerical approach. 

\subsection{Vector mode equations}

The duality relation \eqref{eq:duality} between the monopole and baryonic massless vectors still holds. Particularizing to the background under study and using the ansatz displayed in section \ref{sec:ansatz} one obtains, on the one hand, the first order equation 
\begin{align}\label{eq:a1B8}
\frac{h\,r^6(r-2r_0)^4}{(r-r_0)^4}\alpha'&+16b_Xr^2(\tilde{\alpha}'+b_X\alpha'+Q_k\mathcal{B})+8b_J\frac{r^2(r-2r_0)^2}{(r-r_0)^2}(\hat{\alpha}'+b_J\alpha'-Q_k\mathcal{B})\nonumber\\[2mm]
&-32\sigma_{\smallM}b_J^2b_X-16Q_kb_J(b_J-2b_X)\beta=-\sigma_{\smallB}.
\end{align}
This is used to solve algebraically for $\alpha'$ and substitute in the rest of the equations. On the other hand, the profile of the baryonic massless vector is also fixed by \eqref{eq:duality} in terms of the rest as
\begin{align}\label{eq:A1B8}
\mathcal{A}'&=\sigma_{\smallM} hr^2\left[\frac{8(r-2r_0)^2}{(r-r_0)^2}b_J^2+16b_X^2+\frac{hr^4(r-2r_0)^4}{(r-r_0)^4}\right]\nonumber\\[2mm]
&-\frac{8Q_khr^2(r-2r_0)^2}{(r-r_0)^2}\left[b_J-\frac{2(r-r_0)^2}{(r-2r_0)^2}b_X\right]\beta\nonumber\\[2mm]
&+16b_J\left[2b_X(\hat{\alpha}'-Q_k\mathcal{B})+b_J(\tilde{\alpha}'+Q_k\mathcal{B})+2b_Jb_X\alpha'\right]\,.  
\end{align}
The equations of motion for the massive vectors \eqref{eq:eoma1ha1t} in this geometry read 
\begin{align}\label{eq:a1tB8}
\left[r^2(\tilde{\alpha}'+b_X\alpha'+Q_k\mathcal{B})\right]'&-4(\tilde{\alpha}+\hat{\alpha})-(4a_J-2Q_kb_J+2q_c)\beta'\nonumber\\[2mm]
&-2b_J'(\sigma_{\smallM}b_J-Q_k\beta)=0
\end{align}
and 
\begin{align}\label{eq:a1hB8}
\left[\frac{r^2(r-2r_0)^2}{(r-r_0)^2}(\hat{\alpha}'+b_J\alpha'-Q_k\mathcal{B})\right]'&-8(\tilde{\alpha}+\hat{\alpha})-[8a_J+4Q_k(b_J-b_X)-4q_c]\beta'\nonumber\\[2mm]
&-4b_X'(\sigma_{\smallM}b_J-Q_k\beta)-4b_J'(\sigma_{\smallM}b_X+Q_k\beta)=0\,.
\end{align}
Finally, the equation of motion for the two-form \eqref{eq:eomb2} is translated into two conditions. 
The first condition appears from the component along $*_3\dd a$, and reads 
\begin{align}\label{eq:constraint}
2r^2(\tilde{\alpha}'+b_X\alpha'+Q_k\mathcal{B})&-\frac{r^2(r-2r_0)^2}{(r-r_0)^2}(\hat{\alpha}'+b_J\alpha'-Q_k\mathcal{B})\nonumber\\[2mm]
&-4[Q_k(b_X-2b_J)+2q_c]\beta-2\sigma_{\smallM}b_J(b_J-2b_X)=0\,.
\end{align}
This coincides with (the first integral of) a combination of \eqref{eq:a1tB8} and \eqref{eq:a1hB8} that corresponds to a total derivative. It is the result of that combination of massive vectors giving mass to the two-form, as discussed below \eqref{eq:eomb2}. As a consequence, one can trade one of the second-order equations for the vectors for this condition by solving for $\mathcal{B}$ in terms of the rest. This is equivalent to fixing the gauge of the St\"uckelberg-like symmetry of the massive two-form. Once this is substituted into the equations for the massive vectors, they collapse into a unique second-order equation for the massive combination $\tilde{\alpha}+\hat{\alpha}$. This is also the only combination appearing in \eqref{eq:a1B8}.

The second condition is, instead, dynamical. It arises from the component of \eqref{eq:eomb2} along $\dd a$ and reads
\begin{align}\label{eq:b2B8}
\left[hr^4(r-2r_0)^2\beta'\right]'&-64(\tilde{\alpha}+\hat{\alpha})a_J'-16Q_kb_J(b_J-2b_X)\alpha'-32(2a_J-Q_kb_J+q_c)\tilde{\alpha}'\nonumber\\[2mm]
&-32\left[2a_J+Q_k(b_J-b_X)-q_c\right]\hat{\alpha}'-32Q_k\left[Q_k(b_X-2b_J)+2q_c\right]\mathcal{B}\nonumber\\[2mm]&+\frac{8Q_khr^2(r-2r_0)^2}{(r-r_0)^2}(\sigma_{\smallM} b_J-Q_k\beta)-16Q_khr^2(\sigma_{\smallM} b_X+Q_k\beta)=0
\end{align}
As expected, after substituting the solution to \eqref{eq:constraint}, this equation depends only on the massive vector $\tilde{\alpha}+\hat{\alpha}$. The result of this is a system of two second-order equations for $\tilde{\alpha}+\hat{\alpha}$ and $\beta$, \eqref{eq:a1tB8} or \eqref{eq:a1hB8} together with \eqref{eq:b2B8}, in the understanding that the massless vector profile $\alpha'$ is fixed by \eqref{eq:a1B8} and $\mathcal{B}$ is the solution to \eqref{eq:constraint}. The baryonic vector, and thus the Goldstone, is then obtained from \eqref{eq:A1B8}.

\subsection{Vector mode solutions}

For the numerical analysis, we find it convenient to define the radial coordinate
\begin{equation}
    \uc = \frac{4|Q_k|}{r}\in\left(0,1\right),
\end{equation}
which we use in what follows. Let us denote the remaining massive vector by \mbox{$\gamma_+ \equiv \tilde{\alpha}+\hat{\alpha}$}. Near the boundary, located at $\uc\to0$, the solution takes the following form
\begin{equation}\begin{aligned}\label{eq:expasions_B8_UV}
        \gamma_+ &= 4|Q_k|^ 2\Bigg[\frac{248\, \VV}{2625}\uc^2 + \cdots + \left(\nfour +\frac{472\,  \VV}{2625} \log (\uc)
 \right) \uc^4
  + \cdots
        \Bigg]\,,\\
        \beta &= \frac{8|Q_k|^3}{q_c}\left[\left(
        \frac{25 \sigma_{\smallB}}{3584q_c|Q_k|^2}-\frac{8 \VV}{35}\right)+\cdots+\left(\nfive -\frac{472 \, \VV}{33075}\log(\uc)\right)\uc^5+\cdots\right],\\
	\end{aligned}
\end{equation}
where we have defined $\VV = {q_c^ 2} \sigma_{\smallM}/ (32 |Q_k|^5)$. In these expressions we have made explicit the leading order and the orders at which new parameters appear. We encounter two undetermined dimensionless parameters, $\nfour$ and $\nfive$. In this case, the normalizable solution for $\mathcal{A}$ close to the boundary reads
\begin{equation}
    \mathcal{A} = - q_c^2\, \VV \, \left(\frac{3690496 }{275625}\uc+\frac{4022272 }{3472875}\uc^2+\frac{103790327296 
   }{27348890625}\uc^3+\cdots\right)\,.
\end{equation}

Similarly, we can solve the equations perturbatively around the IR, located at \mbox{$\uc = 1$}. Regularity requires again that $ \sigma_{\smallB} = 0$: as expected, the magnetic field for the baryonic vector is in fact vanishing. Then, the solution reads
\begin{equation}\label{eq:expasions_B8_IR}
	\begin{aligned}
    \gamma_+ &= 4|Q_k|^ 2\left( \nchone (\uc - 1) - \left( \frac{97555}{30622} \nchone - \frac{457}{91866} \nchtwo + \frac{16}{9} \VV \right)(\uc - 1)^2 + \cdots \right)\,,\\
		\beta &= 
		\frac{8|Q_k|^3}{q_c} \left(\nchtwo (\uc - 1) + \left(\frac{31500}{15311} \nchone - \frac{69128}{15311} \nchtwo - \frac{8}{3} \VV \right) (\uc - 1)^2 +\cdots\right)\,.\\
	\end{aligned}
\end{equation}
We end up with $\nchone$ and $\nchtwo$ as the free parameters not determined by the perturbative analysis.

\begin{figure}[t]\centering\includegraphics[width=.49\textwidth]{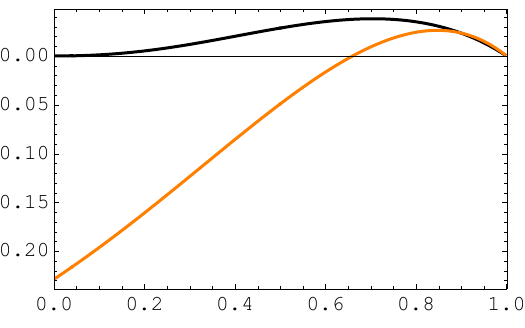}\hfill
	\put(-25,-5){$\uc$}
	\put(-180,135){$\displaystyle \frac{\gamma_+}{4|Q_k|^2}$, $\displaystyle  \frac{q_c\beta}{8|Q_k|^3}$}
    \hfill\includegraphics[width=.49\textwidth]{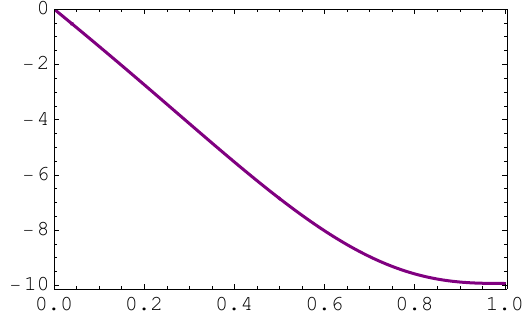}\put(-25,-5){$\uc$}
	\put(-180,130){$\mathcal{A}/q_c^2$}
	\caption{\small Left: numerical solution for $\gamma_+$ (black) and $\beta$ (orange) that interpolates smoothly between the two asymptotic regimes \eqref{eq:expasions_B8_UV} and \eqref{eq:expasions_B8_IR}. 
    Right: corresponding profile of $\mathcal{A}$.
    }\label{fig.Goldstone_B8}
\end{figure}
Having understood how the functions behave at the two boundaries of integration, we connect them numerically. Recall that the equations are linear and homogeneous, meaning that among the five undetermined parameters ($\nfour$, $\nfive$, $\nchone$, $\nchtwo$, and $\VV$), we can fix one, as different solutions are related through rescalings. We set $\VV = 1$, and the remaining four undetermined coefficients match the number of degrees of freedom of our system of equations. The set of values
\begin{equation}
	\VV=1,\quad \nfour = -0.390,\quad \nfive = -0.084.\quad \nchone = -0.306,\quad \nchtwo = -0.377\,,
\end{equation}
is such that the functions are smooth and differentiable everywhere and determine a solution corresponding to the
Goldstone mode. In Fig.~\ref{fig.Goldstone_B8} we plot the result. We conclude that, in this particular model, Chern--Simons terms do not spoil the spontaneous breaking of ${\rm U}(1)_{\smallB}$ as suggested by the presence os a Goldstone boson in the spectrum. 

\newsec{Conclusion and outlook}
\label{sec:discuss}

We have confirmed that baryon symmetry is spontaneously broken in the ${\cal N}=1$, $2+1$-dimensional theories dual to the supergravity solutions constructed in \cite{Cvetic:2001bw,Herzog:2002ss,Faedo:2017fbv}, by explicitly finding the associated Goldstone mode. We have shown this for a confining theory with vanishing Chern-Simons level and for a non-confining theory with a massive spectrum and non-zero Chern-Simons level. We expect that the same result will hold for all the members of the family that do not flow to the IR fixed point,\footnote{We have checked that there is no normalizable and regular mode in the particular background that flows to a CFT in the IR, so there is no Goldstone and therefore ${\rm U}(1)_{\smallB}$ remains unbroken.} as the equations to solve for the vector modes will be the same in qualitatively similar backgrounds. 

This reinforces the similarity of these theories to the Klebanov--Strassler theory \cite{Klebanov:2000hb}, which, as argued in \cite{Aharony:2000pp,Gubser:2004qj}, is a confining ${\cal N}=1$, $3+1$-dimensional theory in the baryonic branch exhibiting a duality cascade. A question is whether there is a moduli space in the three-dimensional theory, which may be present even for this low supersymmetry when time reversal invariance is unbroken \cite{Gaiotto:2018yjh}. We leave a more thorough investigation of this possibility for the future.

A complementary check of the spontaneous breaking of baryon symmetry would be to compute the expectation value of an operator with baryon charge, following analogous calculations in the Klebanov--Strassler theory and some generalizations \cite{Benna:2006ib,Martelli:2007mk,Martelli:2008cm,Gaillard:2013vsa}. A baryon operator is described in the holographic dual by a Euclidean \mbox{D6-brane}, with non-zero flux, wrapped on the entire internal manifold and extended in the radial direction \cite{Aharony:2000pp,Benna:2006ib,Faedo:2023nuc}. Similarly, a dibaryon would correspond to a Euclidean D4-brane wrapped on a four-cycle of the internal geometry. The kappa-symmetry analysis necessary to find the adequate embedding of the branes, along the lines of \cite{Benna:2006ib}, would be interesting but goes beyond the scope of this paper. An added difficulty is that the geometry is not asymptotically AdS, so obtaining some properties of the baryonic operator such as its conformal dimension is not completely straightforward. 

Although we have limited ourselves to zero temperature and baryon charge density, the superfluid phase is expected to extend to a larger region of the phase diagram, bounded by a confinement-deconfinement transition \cite{Faedo:2023nuc}. It would be interesting to follow the evolution of the Goldstone mode throughout this region of the phase diagram, and check whether first and second sound modes emerge, as in the more common realization of spontaneous symmetry breaking by Higgsing the vector fields with a local field in the holographic dual \cite{Herzog:2009md}.

As a final comment, our analysis can also be applied to the unitary $\UU(N_1)\times \UU(N_2)$ gauge theory by simply changing the boundary conditions for the vector fields. In this case there are no Goldstones associated to a possible spontaneous breaking of the global magnetic ${\rm U}(1)_{\smallM}$ symmetry. This can be understood from the ansatz \eqref{eq:ansatz_a1}, where $\sigma_{\smallM}$ is a background magnetic field for monopoles. The regular massless solution is then not normalizable in the unitary theory, as it requires a nonzero $\sigma_{\smallM}$, removing the would-be Goldstones from the spectrum.


\vspace{0.3cm}
\begin{acknowledgments}
The work of A.F. and C.H. is partially supported by the {\em Agencia Estatal de Investigación} and the {\em Ministerio de Ciencia, Innovación y Universidades} through the Spanish grant PID2021-123021NB-I00.
\end{acknowledgments}

\newpage

\bibliographystyle{apsrev4-2}
\bibliography{bibfile.bib}

\end{document}